# Correlated photon pair generation in low-loss double-stripe silicon nitride waveguides

Xiang Zhang, Yanbing Zhang, Chunle Xiong* and Benjamin J. Eggleton

Centre for Ultrahigh bandwidth Devices for Optical Systems (CUDOS), Institute of Photonics and Optical Science (IPOS), School of Physics, University of Sydney, NSW 2006, Australia

*Corresponding author: chunle@physics.usyd.edu.au

**Abstract**. We demonstrate correlated photon pair generation via spontaneous four-wave mixing in a low-loss double-stripe silicon nitride waveguide with a coincidence-to-accidental ratio over 10. The coincidence-to-accidental ratio is limited by spontaneous Raman scattering, which can be mitigated by cooling in the future. This demonstration suggests that this waveguide structure is a potential platform to develop integrated quantum photonic chips for quantum information processing.

**Introduction**:
Photonic quantum technology that harnesses the law of quantum mechanics has intrinsic advantages in several applications, such as quantum computation boosted by superposition [1] and quantum communication secured by entanglement [2]. These applications need a scalable and stable photonic quantum system to generate single or entangled photons, process and analyze quantum information. A promising solution is to develop a CMOS compatible integrated platform which has sufficient nonlinearity for photon generation and low loss for linear optical quantum information processing. Silicon has been used to demonstrate on-chip path entanglement [3], in which silicon rings are used for nonlinear photon generation and silicon-nanowire-based linear circuits are integrated for entanglement analysis. However, a propagation loss of 3 dB/cm in the silicon nanowire is unacceptable in particular circumstances which require long circuitry, for example, time-bin entanglement circuit contains a few centimeters longer arm in unbalanced Mach-Zehnder interferometers (UMZI) for time-bin generation and entanglement analysis [4].

The double-stripe silicon nitride waveguides developed by Lionix have a structure of two vertically separated silicon nitride stripes buried in silica. The novel structure offers low propagation loss (0.2 dB/cm) and tight mode confinement. More importantly, such waveguides can be fabricated using CMOS-compatible technologies at the length of tens of centimeters with high yield [4], which has not been reported in traditional silicon nitride nanowires. These unique features of the double-stripe silicon nitride waveguide has recently enabled the demonstration of compact linear optical circuits consisting of 50 cm long waveguides for on-chip time-bin entanglement [5], However that experiment used a nonlinear photon source from a separate silicon chip. Integrated silicon nitride based nonlinear sources have been demonstrated [6]. Interfacing this nonlinear source with the double-stripe silicon nitride waveguide could be a challenge of fabrication, as these two components have different structures. If this double-stripe waveguide structure has sufficient nonlinearity for photon generation, both on-chip nonlinear sources and integrated linear circuits could be fabricated on the same chip. The classical nonlinear four-wave mixing (FWM) in a double-stripe silicon nitride ring resonator has been reported recently [7], implying that such a waveguide structure could be used for photon generation via spontaneous four-wave mixing (SFWM).

In this paper, we demonstrate correlated photon pair generation in the low-loss double-stripe silicon nitride waveguide. We achieve a pair generation rate of 185 kHz and the corresponding coincidence-to-accidental ratio (CAR) is over 10. The result indicates that the low-loss double-stripe silicon nitride waveguide is a potential platform for integrated quantum photonic chips.

**Principle and experimental setup**:

Correlated photon pairs are generated in the low-loss double-stripe silicon nitride waveguide via SFWM. Figure 1 (a) describes the principle of SFWM: two pump photons at $f_{\text{pump}}$ are annihilated to create a signal photon at higher frequency $f_{\text{signal}} = f_{\text{pump}} + \Delta f$ and an idler photon at lower frequency $f_{\text{idler}} = f_{\text{pump}} - \Delta f$ simultaneously, where $\Delta f$ is the frequency detuning between pump and signal (idler). The generated signal and idler photons are referred to time-correlated photon pairs.

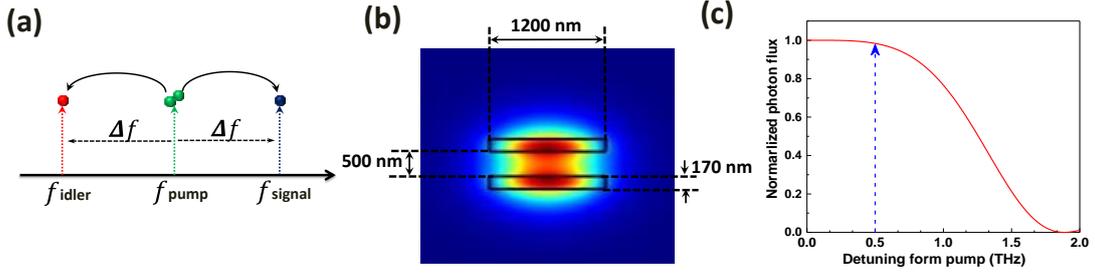

Figure 1 (a) Principle of SFWM. (b) Mode profile in the double-stripe silicon nitride waveguide; the rectangles are the two silicon nitride stripes. (c) Normalized photon flux as a function of frequency detuning.

Figure 1 (b) depicts the cross section of the double-stripe waveguide. Two silicon nitride stripes with a dimension of 170 nm × 1200 nm are vertically separated by 500 nm and buried in silica. Figure 1 (b) also illustrates the mode profile calculated by a finite-element solver (Comsol 5.1). As we can see, the mode is strongly confined in the silicon nitride stripes. Since silicon nitride has much higher nonlinearity than silica [8], the correlated photon pair is expected to be generated in the silicon nitride stripes. Using the actual length ($L$=6.5 cm), the propagation loss ($\alpha = 0.2$ dB/cm), the group velocity dispersion ($\beta_2 = 750 \text{ps}^2/\text{Km}$) [7] and the nonlinearity coefficient ($\gamma = 0.233 \text{ W}^{-1}/\text{m}$) of the waveguide, we obtain the normalized photon flux of double-stripe waveguide through the equation (54) given in Ref. [9]. In figure 1 (c) the full maximum at half width (FWHM) of the photon flux is around 2.4 THz. To avoid pump leakage (investigated in later text) without sacrificing the efficiency of photon pair generation, the frequency detuning between pump and signal (idler) in our experiments is set to 0.5 THz.

Figure 2 shows the experimental setup. Optical pump pulses at 10 ps width are generated from a 50 MHz mode-locked laser (MLL) at the central wavelength of 1555.74 nm. The input power is controlled by a tunable attenuator (ATT). Amplified spontaneous emission noise companioned with pump pulses is eliminated by an AWG which has a channel bandwidth of

50 GHz. A polarization controller (PC) adjusts the pump to be TE polarized before entering the double-stripe silicon nitride waveguide. The waveguide is fiber-pigtailed with polarization maintaining fibers at both ends. The total insertion losses are 4.3 dB including the waveguide-fiber coupling loss (3dB) and the propagation loss (1.3dB). The signal and idler photons are separated by an AWG with an insertion loss of 3 dB before being filtered by a tunable band pass filter (BPF). The BPF has an insertion loss of 2 dB and its bandwidth is 100 GHz. The polarization of generated photons in each channel is adjusted by a PC with an insertion loss of 1 dB. These generated photons are finally detected by the superconducting single photon detectors (SSPDs). The detection efficiency of SSPDs is set to 60% and the dark count rate is around 500 Hz at this efficiency. The collection efficiency from the output of the waveguide to the detector in each channel is 10.5 dB. An electronic delay box (DB) in the idler channel is used to shift the coincidence peak to a positive delay in the histogram of coincidence measurement to avoid the loss of coincidence counts. The resolution of time interval analyser (TIA) is 27 ps.

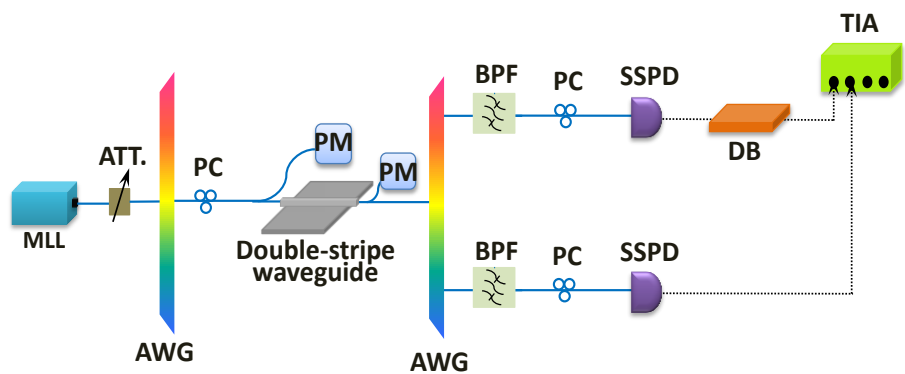

Figure 2 Experimental setup. MLL: Mode locked laser; ATT: attenuator; PC: polarization controller; AWG: arrayed waveguide grating; PM: power meter; BPF: band pass filter; DB: delay box; SSPD: superconducting single photon detector and TIA: time-interval analyzer.

**Experimental results and discussions**:

Figure 3 (a) shows a typical histogram of coincidence measurements. The main peak located at 48 ns represents coincidences, which record the simultaneous detection of signal and idler photons that are generated from the same pump pulses. It indicates the generation of time correlated photon pairs. The coincidence counts contained in the main peak is described by the raw coincidence rate ($C_{\text{raw}}$). Other smaller side peaks with a time interval of 20 ns records the simultaneous detection of photons from different pump pulses, and the count rate obtained from these side peaks represents the accidental rate ($A$). The true coincidence rate $C_{\text{true}} = C_{\text{raw}} - A$ suggests the brightness of a correlated photon pair source, and CAR indicates the strength of photon pair correlations. These two parameters are usually used to characterize the performance of a correlated photon pair source [10].

Figure 3 (b) illustrates the dependency of $C_{\text{true}}$ and CAR on average pump power. The true coincidence rate (diamonds) increases with power, because the efficiency of SFWM scales quadratically with the pump power. The measured true coincidence rate is 1.6 kHz at the pump power of 3.2 mW. Taking into accounts the collection efficiency, the 1.6 kHz true coincidence rate corresponds to a correlated photon pair generation rate of 185 kHz in the waveguide. CAR (squares) reduced from 16 to 10.6 with the increase of the power due to the generation of multi-pair via SFWM. When the power is less than 3.2mW, a CAR over 10 is achieved. This suggests the correlated photon pair source can be used for quantum key distribution with 95% error free [11]

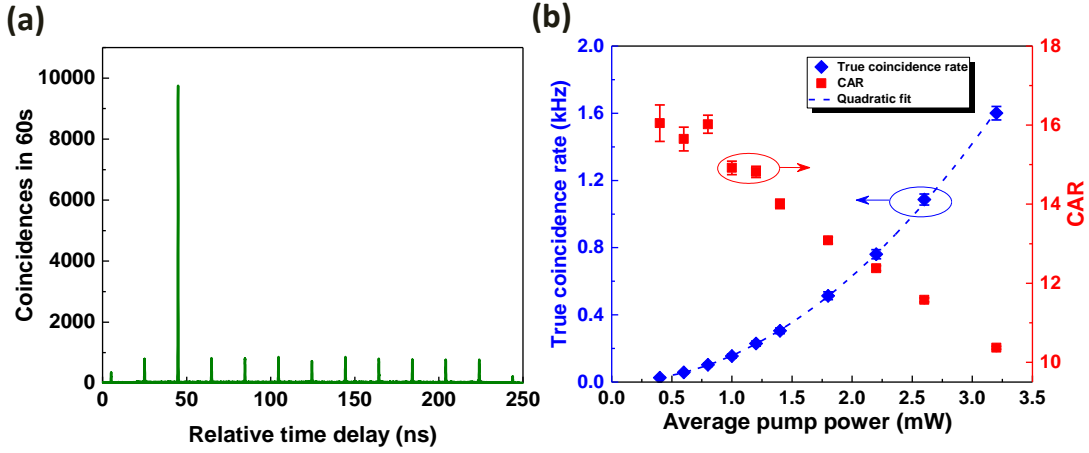

Figure 3 (a) A typical histogram of coincidence measurements at the input average pump power of 3.2 mW. (b) True coincidence rate (diamonds) and CAR (squares) as a function of average pump power; dashed line is a quadratic fit. Error bars are calculated by Poissonian distribution.

We analyze the CAR without other noise except multi-pairs and we find the overall value of CAR in figure 3 (b) is far below our expectation. We assume there is only the multi-pair noise and $CAR \approx 1/\mu$. Here, $\mu = C_{\text{true}}/\eta_i\eta_s R$ is average number of photon pairs per pulse, $\eta_i$ and $\eta_s$ are the collection efficiencies of idler and signal photons respectively, and $R$ is the laser repetition rate [12]. Using the measured true coincidence rate at the input power of 3.2 mw, we find CAR should be over 270. However, in figure 3 (b), CAR is far below this expectation. This significant discrepancy suggests that there are other noise sources contribute to this entire reduction, such as pump leakage and spontaneous Raman scattering (SpRS), which will be discussed in the following paragraphs.

We first investigate the pump leakage via coincidence measurements at different pump-idler/signal frequency detunings. The pump leakage results from the spectrum overlap between the pump channel and the selected signal/idler channel. If pump leakage contributes to the majority part of noise photons, the accidental peaks should be similar to the main peak, since the correlation between the detections in signal and idler channel is mainly determined by pump photons rather than correlated photon pairs. The spectrum overlap decreases significantly with the increase of frequency detuning. In figure 4(a), we show the histogram of coincidence measurements with three different frequency detunings at the same average pump power of 3.2 mW. To discriminate the difference between the main peak and the side peaks at each frequency detuning, we intentionally introduce different delays at each frequency detuning via the DB. When the frequency detuning $\Delta f = 400$ GHz, the accidental peaks are similar to the main peak located at 40 ns. In the case of $\Delta f = 500$ GHz (600 GHz), the accidental peaks are quite trivial compared with the main peak, located at 48 ns (52 ns). Note the decrease of the main peak at $\Delta f = 600$ GHz compared to $\Delta f = 500$ GHz comes from the phase mismatch of SFWM (see figure 1 (c)). Therefore, to minimize the pump leakage, we choose a frequency detuning of 500 GHz.

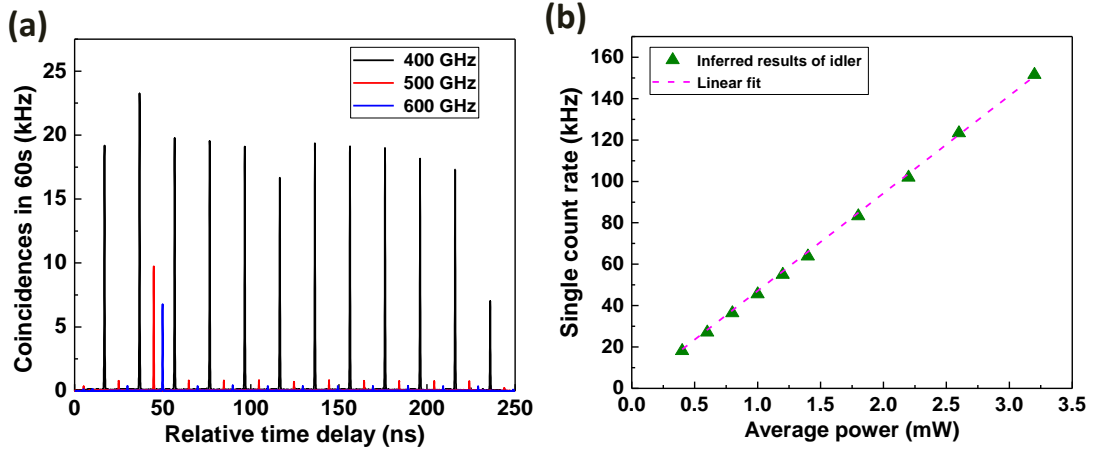

Figure 4 (a) Histogram of coincidence measurements at three different frequency detunings. (b) Inferred single count rate generated from SpRS. Triangles are the inferred single counts rate of idler. Dashed line is a linear fit.

Now, we investigate the noise induced by SpRS. Silicon nitride has a broadband Raman emission that includes frequency shift <44.22 THz and 93.46-114.76 THz [13]. The measured single counts include photons generated via both SFWM and SpRS. The photon generation via SFWM scales quadratically with the pump power while SpRS scales linearly. The coincidence analysis in figure 3 (b) has already confirmed the quadratic dependence. Here, we use the idler counts to explain the SpRS-induced noise, since the signal channel has similar single counts. The inferred single counts ($C_{i\_SpRS}$) from SpRS is calculated by $C_{i\_SpRS} = C_{i\_total} - C_{i_{SFWM}} - D$. Here, $C_{i\_total}$ is the total counts detected in idler channels, $C_{i\_SFWM}$ is the single counts generated from SFWM and $D$ represent the dark counts. $C_{i\_SFWM}$ is analyzed by $\mu\eta_i R$. Figure 4 (b) illustrates the inferred results of idler (triangles) as a function of the input pump power. The linear fit (dashed line) indicates that the noise photons are most likely contributed by SpRS, as SpRS is linearly dependent on the pump power.

To further confirm that correlated photon pairs are generated in the 6.5 cm long waveguide rather than in the 6 m long connection silica fibers. We replace the double-stripe silicon nitride waveguide and the pigtailed polarization maintenance fibers by a tunable ATT and two fibers with similar length to repeat the measurement. The total loss of the replacement has been adjusted to 4.3 dB to simulate the waveguide loss. At the same average pump power, the true coincidence rate in silica fibers is less than 4% of the true coincidence rate in the double-stripe silicon nitride waveguide. This confirms that correlated photon pairs are indeed generated in the silicon nitride waveguide.

In the future, we can further improve both the source brightness and CAR. For example, we could design a ring resonator as an ultra-compact nonlinear source component. Because of the high Q factor, correlated photon pair generation in a ring resonator is much more efficient than in a straight waveguide with the same input pump power. Furthermore, the required bandwidth of an on-chip filter is not stringent due to the extremely narrow line width of the ring resonator which has already guaranteed the purity of correlated photon pairs. On the other hand, the noise photons from SpRS could be reduced significantly by cooling. More specifically, the Raman noise could be reduced by 70.6%, using the SpRS equation in Ref. [14], if the sample is cooled down to the temperature of liquid nitrogen.

**Conclusion**:

In conclusion, we have experimentally demonstrated correlated photon pair generation in a low-loss double-stripe silicon nitride waveguide via SFWM. We have achieved a CAR over 10 when the power is less than 3.2 mW. We find that the CAR is mainly limited by SpRS which could potentially be mitigated by cooling. This work implies that the low-loss double-stripe silicon nitride waveguide is a promising structure for integrating both nonlinear sources and linear circuits in a monolithic photonic chip. It takes us a step further towards integrated quantum photonic chips for various applications.

**Acknowledgement**:

The authors would like to thank Lionix for chip fabrication and Jiakun He for helpful discussion. The work is supported by the Centre of Excellence (CUDOS, CE110001018), Laureate Fellowship (FL120100029) and Discovery Early Career Researcher Award (DE120100226) programs of the Australian Research Council (ARC).